\documentclass[usenatbib]{mnras}

\usepackage[utf8]{inputenc}
\usepackage[T1]{fontenc}
\usepackage{graphicx}	
\usepackage{amsmath}	
\usepackage{amssymb}	
\usepackage{multicol}   
\usepackage{bm}		    
\usepackage{pdflscape}
\usepackage{hyperref}
\usepackage{xcolor}

\title[Quenching and Galaxy Demographics]{\textit{Quenching and Galaxy Demographics}}
\author[L. F. de la Bella, et al.]{L.~F.~de la Bella$^{1,2}$\thanks{E-mail: lucia.fonseca-de-la-bella@port.ac.uk}, A.~Amara$^{1}$, S.~Birrer$^{3,4}$,
W.~G.~Hartley$^{5}$ and
P.~Sudek$^{1}$
\\
$^{1}$Institute of Cosmology and Gravitation, University of Portsmouth, Portsmouth, P01 3FX, UK.\\
$^{2}$Department of Physics \& Astronomy, The University of Manchester, Manchester M13 9PL, UK.\\
$^{3}$Kavli Institute for Particle Astrophysics and Cosmology and Department of Physics, Stanford University, Stanford, CA 94305, USA.\\
$^{4}$SLAC National Accelerator Laboratory, Menlo Park, CA, 94025.\\
$^{5}$Department of Astronomy, University of Geneva, ch. d’Ecogia 16, CH-1290 Versoix, Switzerland
}

\date{Accepted XXX. Received YYY; in original form ZZZ}

\pubyear{2021}

\begin{document}
\label{firstpage}
\pagerange{\pageref{firstpage}--\pageref{lastpage}}
\maketitle

\begin{abstract}
The open-data revolution in astronomy is forcing the community to develop sophisticated analysis methods that heavily rely on realistic simulations. The  phenomenology of the evolution of galaxy demographics can be described by a set of continuity equations invoking two quenching mechanisms: mass quenching and satellite quenching. The combination of these two mechanisms produces a double Schechter function for the quiescent population, as is observed in the low-redshift universe.  In this paper we consider these quenching mechanisms, explicitly including satellite galaxies, and add the exact time evolution of the star-forming population. These new features complete the current versions of these continuity equations, and are essential for the realistic simulations required in modern extra-galactic astrophysics.  We derive the analytical relation between the quiescent and the active populations, reducing considerably the parameter space for the simulation. In addition, we derive the analytical time dependence of the amplitude of the Schechter function. Finally, we validate our results against the SDSS DR7 galaxy sample. The model will be implemented in the \verb!SkyPy! library and the main plots sonified using \verb!STRAUSS!.
\end{abstract}


\section{Introduction}
\label{sec:introduction}

Astronomy is currently living an open-data revolution led by legacy surveys such as Euclid \citep{Euclid2011}, the Rubin Observatory Legacy Survey of Space and Time \citep{LSST2019}, Planck \citep{Planck2020} and the Laser Interferometer Gravitational Wave Observatory \citep{LIGO2015}. The main barrier to research in such a revolution is access to increasingly sophisticated analysis methods. For example, forward modelling and machine learning have emerged as important techniques for the next generation of surveys. Both of these techniques depend heavily on realistic simulations. In this context, accurate representations of the galaxy populations in the simulation and analysis of the ongoing and future large-scale cosmology experiments are essential.

In this paper, we study the phenomenology of the evolution of galaxy demographics and implement an accurate prescription based on a quenching model to generate galaxy catalogues. Traditionally the galaxy mass distribution is described by the Schechter function \citep{Schechter_1976}, and two main galaxy populations are distinguished: active and quiescent. The active or star-forming population is composed of galaxies actively forming stars, increasing their stellar mass. Conversely, the quiescent population is made of quenched galaxies which do not create stars (or do so very slowly). Typically two quenching mechanisms transform star-forming galaxies into quiescent objects: mass quenching and satellite quenching (see for example \cite{Peng_2010}). Satellite quenching usually occurs when a subhalo (and its galaxy) enters a denser region of space, e.g. when falling into a parent halo. Physically this phenomena has been related to strangulation \citep{Larson_1980, Balogh_2000} and ram pressure stripping \citep{GunnGott_1972}. Likewise, the primary cause of mass quenching is believed to be feedback from active galactic nuclei or supernovae \citep{Fabian_2012}. However, there exist other potential interpretations that consider these two mechanisms as different manifestations of a common group quenching \citep{Knobel_2015}.

In the astrophysical literature, quiescent galaxy samples at low redshift are commonly described by a double Schechter function \citep{LiWhite_2009, Peng_2010, Pozzetti_2010, Baldry_2012, Ilbert_2013, Muzzin_2013, Birrer_2014}, which is the addition of two single Schechter functions. Authors such as \cite{Peng_2010, Peng_2012} empirically connect the star-forming mass function via the quenching phenomena or ceasing of star formation. Typically, these findings are drawn from observational data, employing different methods such as the classical $1/V_{max}$ approach \citep{Schmidt_1968}, parametric maximum likelihood methods \citep{Sandage_1979} or  non-parametric  step-wise-maximum likelihood techniques \citep{Efstathiou_1988}. Alternatively, similar conclusions are drawn from more physical approaches, i.e. \cite{Birrer_2014}.

In this work we derive the continuity equations describing the rate of population change with fixed mass and environment. The model distinguishes between satellite galaxies and central galaxies and considers the probability of being satellite quenched, mass quenched and growing in stellar mass. 
By solving the equations analytically we identify a double Schechter function for the quiescent galaxies and validate our model against the best-fit SDSS DR7 sample from \cite{Weigel_2016}.

\begin{figure}
\centering
\includegraphics[width=0.35\textwidth]{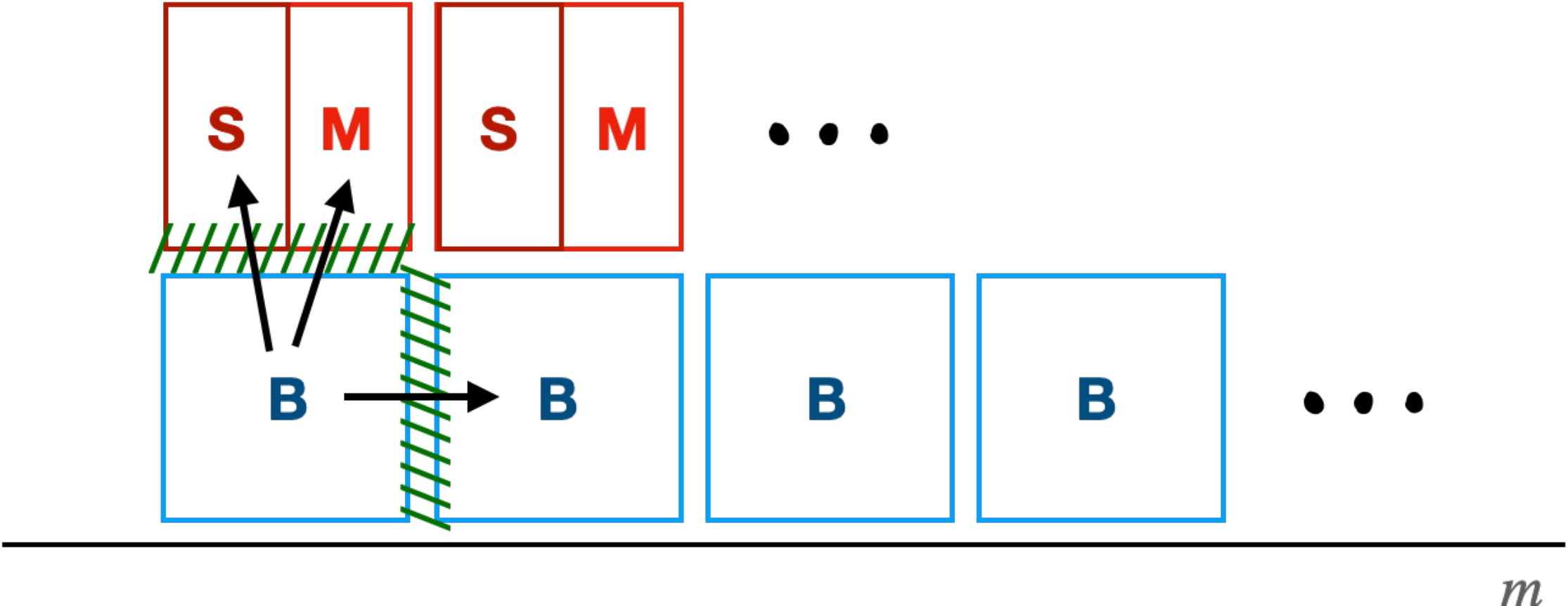}
\caption{This diagram represents the growth of the blue population and quiescent populations described by Peng et al. \citep{Peng_2010}. The axis represents the infinitesimal mass bins, $B$ the active population, $S$ the satellite-quenched and $M$ the mass-quenched galaxies. In every bin, the blue galaxies can come out through the boundaries (hatched green lines) by growing or becoming quiescent.}
\label{fig:quenching}
\end{figure}

A similar model was originally described by \cite{Peng_2010} (c.f. Figure \ref{fig:quenching}). The authors derived a set of continuity equations for the number of blue galaxies lost per unit time in a given infinitesimal mass bin. In this picture they distinguish between three different mechanisms: \textit{a)} star formation, the galaxy grows in mass and moves to the next bin, and \textit{b)} satellite and/or mass quenching, leaving the galaxy with fixed mass and moving to the quiescent sample. However, it is not clear how their model would account for satellite galaxies and the incoming blue galaxies from less massive bins due to growth. In that sense, their model is incomplete and cannot be implemented within simulation pipelines without additional specification and constraints. In this paper we include satellite galaxies explicitly and present these evolution processes as a continuous Markov chain, accounting correctly for the growth in stellar mass.

Additionally, our model will be included in the \verb!skypy.galaxies! module of the \verb!SkyPy! library\footnote{\url{https://github.com/skypyproject/skypy.git}} after the publication of this manuscript. \verb!SkyPy! ~\citep{Amara_2021, skypy_collaboration_2020_3755531} is an open-source Python package for simulating the astrophysical sky. It comprises a library of astrophysical models and a command -line script to run end-to-end simulations. With the implementation of our galaxy demographics model, the user can also draw active and quiescent populations. The new functionality and pipeline to reproduce our results will appear in the examples page of the \verb!SkyPy! documentation\footnote{\url{https://skypy.readthedocs.io/en/stable/}}.

In this paper we present the quenching model in Section \ref{sec:the_quenching_model} and identify the solutions of the continuity equations with the star-forming (single) and quiescent (double) Schechter functions. In Section \ref{sec:schechter_mass_function} we show how the Schechter parameters for quiescent galaxies are related to the star-forming parameters, identifying the satellite-quenched galaxies as a subset of the blue sample. In addition, we present the time evolution of the amplitude of the star-forming Schechter function. Then we validate our results using a fitting curve to the SDSS DR7 sample in Section \ref{sec:validation}.
Finally we conclude and introduce our future lines of work in Section \ref{sec:conclusions}.

\section{The quenching model}
\label{sec:the_quenching_model}

In this section we describe the quenching phenomena associated with an increase in stellar mass and an increase in the density of the environment. 
The mass function describing the number density distribution of galaxies as a function of stellar mass is given by the aforementioned Schechter distribution \citep{Schechter_1976}
\begin{equation}\label{eq:schechter}
    \phi(m, t) = \phi_{*}(t) \left(\frac{m}{m_{*}} \right)^{\alpha} e^{- \frac{m}{m_{*}}} \; ,
\end{equation}
where $\phi_{*}$ is the amplitude of the Schechter function, $\alpha$ is the faint-end slope parameter and $m_{*}$ is the characteristic mass in units of solar masses.  Different values of this set of parameters will describe both the star-forming and the quiescent populations.

\begin{itemize}
    \item \textbf{Active galaxies}. Active or blue galaxies correspond to the population that actively forms stellar mass. In the cosmological model, dark matter halos are formed and grow by acquiring smaller halos. These halos host galaxies which grow in stellar mass. As they become more massive, some of these central galaxies start feeling the gravitational pull from other larger objects and turn into satellite galaxies. Therefore there exists a probability of a central galaxy becoming a satellite galaxy, $\eta_{\rm sat}$. This probability depends on mass, i.e. more massive galaxies will tend to remain central whereas smaller galaxies will become satellites. This probability is related to the fraction of satellite galaxies 
    \begin{equation}\label{eq:fsat}
        f_{\rm sat} \equiv \frac{n_{\rm sat}}{n_{\rm total}} 
    \end{equation}
    with $n_{\rm sat}$ the number density of satellite galaxies and $n_{\rm total}$ the number density of the total galaxy sample of a given mass. 
    The increase in mass is driven by the star formation rate of the long-lived stellar population 
    \begin{equation}\label{eq:sfr}
         SFR \equiv \frac{dm}{dt}
    \end{equation}
    with the specific star-formation rate defined as $sSFR \equiv SFR/ m$.

    \item \textbf{Mass quenching}.  Mass quenching is the cessation of star formation when a blue galaxy reaches a critical mass. Let us consider a galaxy within a constantly accreting halo. At the beginning, both galaxy and halo are growing together \citep{Birrer_2014}. Eventually, the galaxy is quenched and stops growing, fixing its stellar mass; even though the halo continuous to grow. Note here that galaxy-galaxy mergers will eventually increase the galaxy's stellar mass post-quenching.
    
    This process is characterised by the mass-quenching rate which is the probability of a galaxy being mass quenched per unit time. This transformation rate represents the fraction of active galaxies that are mass-quenched, $f_m$. According to \cite{Peng_2010}, the mass-quenching law is given by
    \begin{equation}\label{eq:massq}
        \eta_m = \mu SFR
    \end{equation}
    with $\mu = m_{*}^{-1}$ and $SFR$ the star formation rate.
    
    Equation \eqref{eq:massq} is valid for all masses and environments and at all epochs. The quenching rate  has units of inverse $Gyr$ and can be interpreted as the time that a blue galaxy statistically awaits to be mass-quenched.
    
    In terms of the Schechter function, we will show that this population presents the same characteristic mass, $m_{*}$, as the blue population but with a faint-end slope parameter which differs from the blue population by one unit. This was originally proved by \cite{Peng_2010}.\\
    
    \item \textbf{Satellite quenching}. Satellite quenching is the cessation of star formation potentially due to an increase in the density of the environment. Note that the physical cause and time scales of satellite quenching are still the subject of ongoing debate.  As a simplified picture, we assign an instantaneous probability of the in-falling satellite to be quenched \citep{Birrer_2014}. Finally, the blue survivals are subjected to growth and mass quenching as they become massive.
    
    This phenomenon is characterised by the satellite-quenching rate which is the probability of a blue satellite galaxy being satellite-quenched per unit time. This transformation rate represents the fraction of active galaxies, hosted by subhalos, that are satellite-quenched, $f_{\rho}$. According to \cite{Peng_2010} this quenching rate is given by
    \begin{equation}\label{eq:envq}
        \eta_{\rho} = \frac{1}{1 - \epsilon_{\rho}}\frac{\partial \epsilon _{\rho}}{\partial \log \rho} \frac{\partial \log \rho}{\partial t}
    \end{equation}
    with $\epsilon_{\rho}$ the quenching efficiency that depends on the comoving density of the environment, $\rho$. Again, the quenching rate  has units of $Gyr^{-1}$ and can be interpreted as the time that a blue galaxy statistically awaits to be satellite-quenched.
    In principle, the fraction of galaxies turning quiescent through satellite quenching should be redshift dependent, although slowly
    \begin{equation}\label{eq:frho_etarho}
        \partial f_{\rho}/ \partial t = \eta_{\rho} (1 - f_{\rho})\, . 
    \end{equation}
    For simplicity we consider this a constant,  $0< f_{\rho} < 1$ \citep{Birrer_2014}. For a more realistic scenario refer to \cite{Hartley_2013}.
    
    In terms of the Schechter function, we will show that this quiescent population is essentially a subset of the blue population. They have the same shape, $\alpha$, same characteristic mass, $m_{*}$, but lower amplitude \citep{Peng_2010}.
\end{itemize}
Besides quenching there exist further complex phenomena in the picture of galaxy evolution such as galaxy-galaxy merging. The merging of galaxies would impact the massive end in the quiescent sample distribution, increasing the number density of massive red galaxies. The modelling of such processes will be the scope of future work.

\subsection{Continuity Equations for Galaxy Demographics}
\label{sec:continuity_equations}

\begin{figure}
\centering
\includegraphics[width=0.3\textwidth]{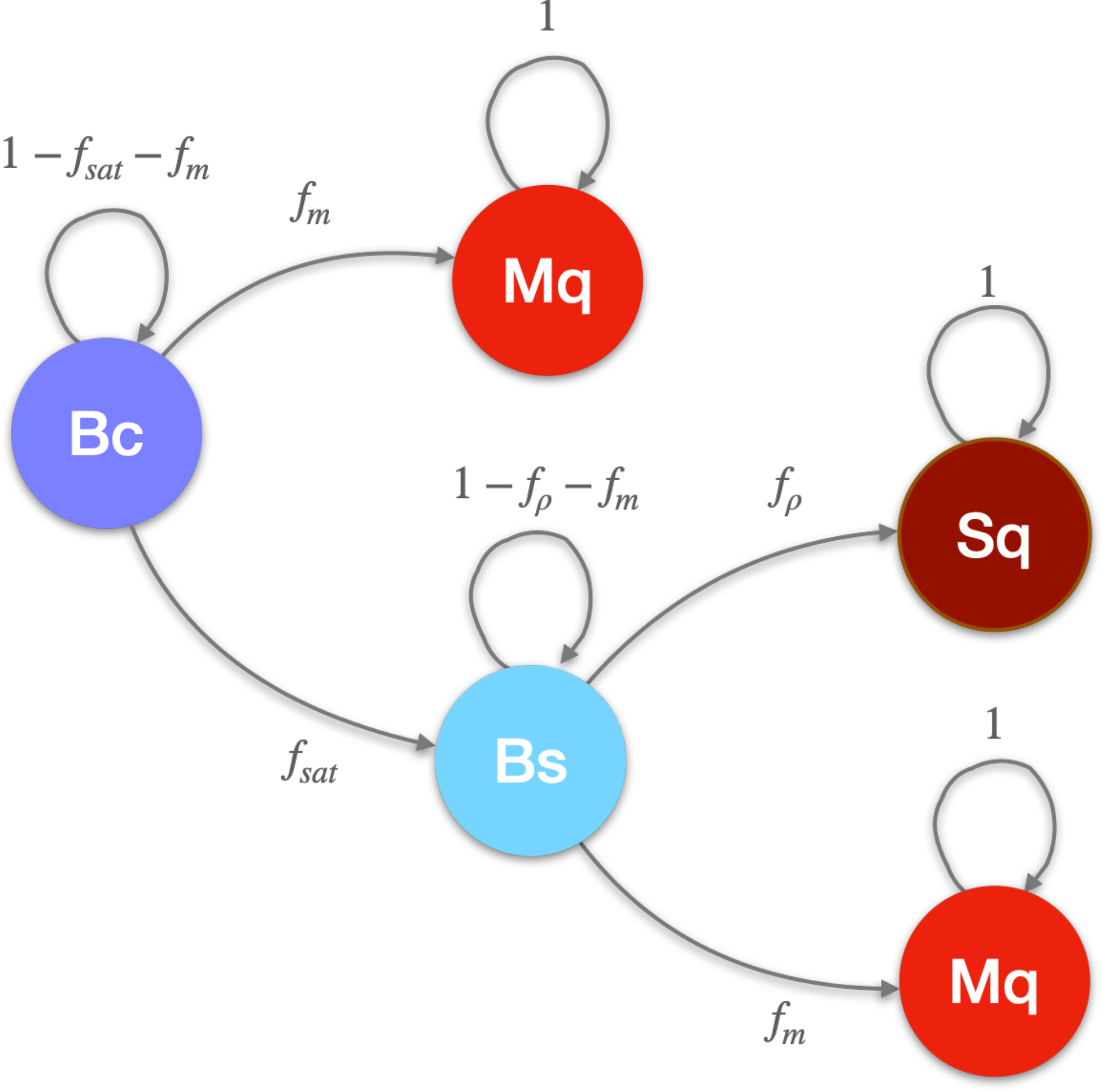}
\caption{This is the Markov chain representing the population change for an infinitesimal time. $B$ represents active galaxies, $B_c$ central galaxies, $B_s$ satellite galaxies, $M_q$ mass quenched galaxies and $S_q$ satellite quenched galaxies. Please refer to text for a detailed explanation.}
\label{fig:markov}
\end{figure}

In this work we interpret the galaxy evolution as a Markov process  \citep{Birrer_2014}. In Figure \ref{fig:markov} the chain starts with a blue central galaxy, $B_c$, with a probability of becoming a satellite galaxy, $B_s$. The central galaxy could remain active and grow (determined by the star-formation rate) or could be too massive with a probability of being mass-quenched, $M_q$. If it becomes a satellite galaxy, it could be eventually satellite quenched, $S_q$. Otherwise, the satellite galaxy could also be mass-quenched or remain active and grow. In our model we consider that once a galaxy had been quenched there is no way to become active again.  This simple prescription not only has the power to show the connection between the quenching phenomena and the different populations, but also allows  us to get individual star-formation histories, including for quenched galaxies, which is not possible with the \cite{Peng_2010} formalism alone. We will show at the end of the next section how this translates into a double Schechter function for the quiescent galaxies and how they relate to the active population.

From the above description we derive the equations that govern galaxy evolution
\begin{equation}\label{eq:evolution}
\begin{split}
   \left.\frac{d B_c}{dt}\right\vert_{m, \rho} & = \alpha sSFR\, B_c - \eta_m B_c - \eta_{\rm sat} B_c\\
    \left.\frac{d B_s}{dt}\right\vert_{m, \rho} & = \alpha sSFR\, B_s - \eta_m B_s - \eta_{\rho} B_s + \eta_{\rm sat} B_c \\
    \left.\frac{d M_q}{dt}\right\vert_{m, \rho} & = \eta_m B_c + \eta_m B_s \\
    \left.\frac{d S_q}{dt}\right\vert_{m, \rho} & = \eta_{\rho} B_s
\end{split}
\end{equation}
where $B_c$ denotes the number density of central galaxies, $B_s$ satellite galaxies, $M_q$ mass-quenched galaxies and $S_q$ satellite-quenched galaxies. Note the increase of the number density of active galaxies due to ongoing star formation is characterised by the logarithmic slope of the mass function \eqref{eq:schechter}, $\alpha \equiv d \log \phi / d \log m$, and the sSFR-mass relation, $sSFR$.\\

In order to solve equations \eqref{eq:evolution}  analytically we consider the following list of assumptions: \textit{a)} $\alpha$ and $m_{*}$ are constant, \textit{b)} the evolution of the galaxy distribution with time is very slow,  \textit{c)} the fraction of satellite galaxies $f_{\rm sat}$ evolves slowly with time and  $f_{\rho}$ is constant, and \textit{d)} the initial conditions are set at a time where stellar mass was very low $m_0 << m_{*}$, as well as the environmental density. This choice implies that the initial blue sample was only composed of central galaxies, $B_0 = B_{c0}$ ($B_{s0} = 0$), there existed no quiescent galaxies, $M_{q0} = S_{q0} = 0$, and therefore $f_{sat0} = f_{\rho 0} = 0$.

The solutions for the active population read
\begin{equation}\label{eq:nblue}
    \begin{split}
        B_c(m, t) & = B_{c*}(t) \left( \frac{m}{m_{*}}\right) ^{\alpha} e^{- \frac{m}{m_{*}}} \\
        B_s(m, t) & = B_{s*}(t) \left( \frac{m}{m_{*}}\right) ^{\alpha} e^{- \frac{m}{m_{*}}}
    \end{split}
\end{equation}
with amplitudes
\begin{equation}\label{eq:nblue_amplitude}
    \begin{split}
        B_{c*}(t) & = B_0 \left( \frac{m_{*}}{m_0}\right) ^{\alpha} e^{ \frac{m_0}{m_{*}}} e^{- \int_{t_0}^t \eta_{\rm sat} dt'} \\
        B_{s*}(t) & = B_{c*}(t) \int_{t_0}^t \eta_{\rm sat} dt' \, .
    \end{split}
\end{equation}
Note that when setting $t$ to the initial time we retrieve the expected results $B_c(m, t_0) = B_0$ and $B_s(m, t_0) = 0$.

And the quiescent galaxies
\begin{equation}\label{eq:nred}
    \begin{split}
        M_q(m, t) & \simeq M_{q*}(t) \left( \frac{m}{m_{*}}\right) ^{\alpha + 1} e^{- \frac{m}{m_{*}}} \\
        S_q(m, t) & = S_{q*}(t) \left( \frac{m}{m_{*}}\right) ^{\alpha} e^{- \frac{m}{m_{*}}}
    \end{split}
\end{equation}
where we have made explicit use of $ m_0 << m_{*}$ and that the time evolution of the blue galaxy distribution is very slow. The amplitudes read
\begin{equation}\label{eq:nred_amplitude}
    \begin{split}
        M_{q*}(t) & = B_{c*}(t) + B_{s*}(t) \\
        S_{q*}(t) & =  B_{s*}(t) \int_{t_0}^t \eta_{\rho} dt'
    \end{split}
\end{equation}
Note that indeed $M_q(m, t_0) \simeq 0$ and $S_q(m, t_0) = 0$.

In summary, we showed in this section how the galaxy demographics can be presented as a Markov chain and described as a set of continuity equations that can be solved analytically.

\section{The Quiescent Schechter Mass Functions}
\label{sec:schechter_mass_function}

In this section, we show the relation between the quiescent Schechter parameters and the properties of the blue population. At the end of the section we study the explicit time dependence of the amplitude of the mass function.

\subsection{Reduction of the parameter space}

By inspection of the equations above \eqref{eq:nblue}, we deduce that evidently the star-forming population follows a Schechter mass function \eqref{eq:schechter} 
\begin{equation}\label{eq:schechter_blue}
    \phi_{b}(m, t) = B_c(m, t) + B_s(m, t)
\end{equation}
with $\phi_{*b}(t) = B_{c*}(t) + B_{s*}(t)$ given by \eqref{eq:nblue_amplitude}, $\alpha_{b} = \alpha$ and $m_{*b} = m_{*}$.

From equations \eqref{eq:nred} and \eqref{eq:nred_amplitude}, we demonstrate how the mass-quenched population has the same amplitude and characteristic mass than the active population but a different faint-end slope
\begin{equation}\label{eq:massq_params}
\begin{split}
    \alpha_{m} & = \alpha_{b} + 1 \\
    m_{*m} & = m_{*b}\\
    \phi_{*m}(t) &  \simeq \phi_{*b}(t)\; .
\end{split}
\end{equation}
Likewise, the  satellite-quenched galaxies are clearly a subclass of the active population, although with lower amplitude
\begin{equation}\label{eq:satq_params}
\begin{split}
    \alpha_{\rho} & = \alpha_{b}\\
    m_{*\rho} & = m_{*b}\\
    \phi_{*\rho}(t) & = F_{\rho} \phi_{*bs}(t)
\end{split}
\end{equation}
with $\phi_{*bs}(t) = B_{s*}(t)$ in \eqref{eq:nblue_amplitude} and  $F_{\rho} \equiv \int_{t_0}^t \eta_{\rho} dt' = \mathrm{ln} (1 / (1- f_{\rho}))$, using equation \eqref{eq:frho_etarho}.

These relations reduce the parameter space from nine to five parameters
\begin{equation}\label{eq:parameter_space}
    \begin{Bmatrix}
        \phi_{*b} & \alpha_b & m_{*b}\\
        \phi_{*m} & \alpha_m  & m_{*m}\\
        \phi_{*\rho} & \alpha_{\rho} & m_{*\rho} 
    \end{Bmatrix}
    \longrightarrow 
    \begin{Bmatrix}
        \phi_{*b} & \alpha_b & m_{*b}\\
        f_{\rho} & f_{\rm sat} & 
    \end{Bmatrix}
\end{equation}
with the possibility of reducing it to four parameters $\{\phi_{*b}, \alpha_b, m_{*b}, f_{\rho}\}$ if the separation between satellites and central galaxies is known.

\subsection{Time evolution of the Schechter Function}

The time dependence of the galaxy population is crucial to track the galaxy evolution throughout cosmic time. In the literature there exist many empirical expressions and parametrisations of such time dependence. One example of the parametrisation of the time evolution of the amplitude of the Schechter function is the model used by \cite{Herbel}

\begin{equation}\label{eq:herbel}
    \phi_{*}(z) = b e^{az}
\end{equation}
where $z$ is redshift and $a$ and $b$ free fitting parameters.

As a novelty, we need not perform any parametrisation since we obtain the exact analytical solutions. We can re-write the amplitude of the active population \eqref{eq:schechter_blue} as a function of redshift
\begin{equation}\label{eq:amplitude_z}
    \phi_{*b}(z) = A e^{f(z)}
\end{equation}
where $A$ is a combination of prefactors in equations \eqref{eq:nblue_amplitude} and $f(z)$ is the argument of the exponential $ e^{- \int_{t_0}^t \eta_{\rm sat} dt'}$ written in terms of redshift. One can clearly observe that had we not any information about $f(z)$, we would perform a polynomial expansion as a first approach, retrieving the parametrisation in equation \eqref{eq:herbel} \citep{Herbel}. \\

In this section we showed how these distributions correspond to Schechter functions. We also justified the appearance of the double Schechter function  for the quiescent populations, demonstrating that the satellite-quenched galaxies are indeed a subset of the blue galaxies and that the mass-quenched galaxies have a different faint-end slope parameter. This connection between the quenching phenomena and the galaxy populations allowed us to reduce the parameter space and derive the analytical time dependence of the amplitude of the Schechter function for the first time in the literature.

\section{Validation}
\label{sec:validation}

In this section we validate our model using the results from the best fit to SDSS DR7 data in \cite{Weigel_2016}. The authors present a comprehensive method to determine stellar mass functions and apply it to samples in the local universe, in particular to SDSS DR7 data in the redshift range from $0.02$ to $0.06$. Note that we are only comparing our results to a fit. However, this is a reasonable procedure and expect the same outcome when directly matching to real data, since the fit is a sufficient representation to the current data.

\begin{figure*}
\label{fig:weigel}
\centering
\includegraphics[width=0.9\textwidth]{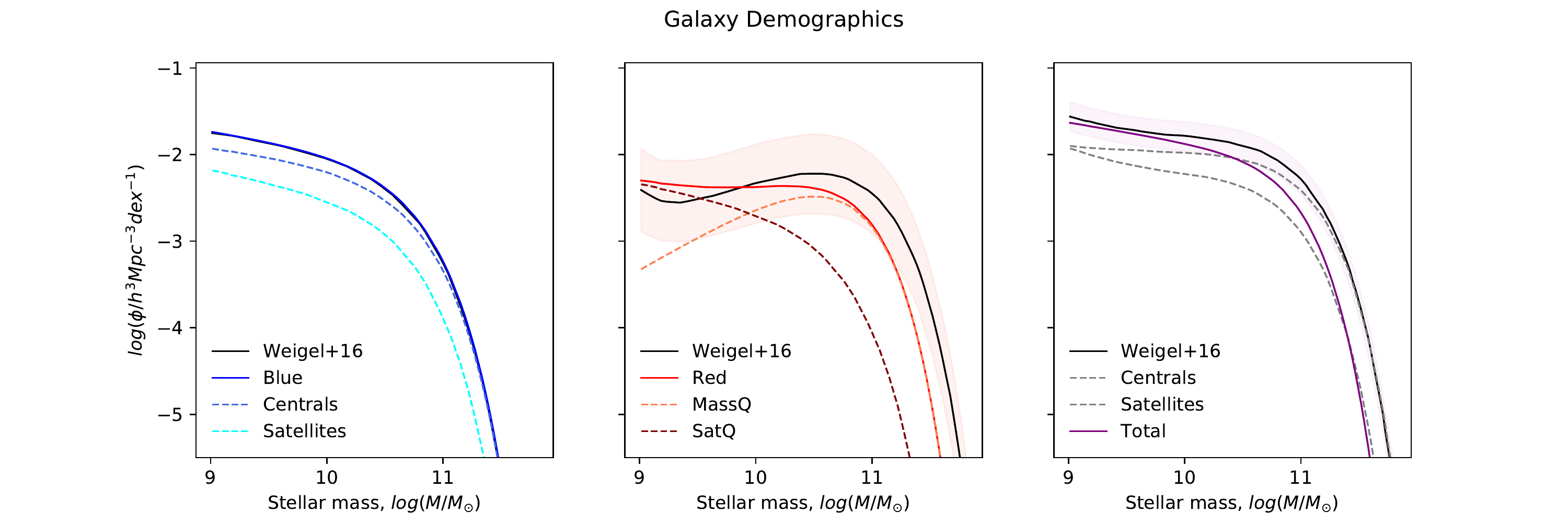}
\caption{
Model from equations \eqref{eq:nblue} and \eqref{eq:nred} compared to Weigel et al. \citep{Weigel_2016}.  From left to right, we plot the different populations: active galaxies, quiescent galaxies and the entire sample. The solid black lines corresponds to the best fitting model from \citep{Weigel_2016}, and the dashed grey lines in the right panel represent the central and the satellite population of the entire sample. On the left, the dashed blue lines correspond to the survival central and satellite galaxies \eqref{eq:nblue}, whereas the solid blue line corresponds to the entire blue sample \eqref{eq:schechter_blue}. On the middle, the dashed lines represent the mass-quenched and satellite-quenched galaxy populations \eqref{eq:nred} and the solid red line corresponds to the total quiescent population. On the right panel, the purple line is the total sample using our model. Our simple model successfully produces two quiescent populations linked to different quenching processes, justifying the double Schechter function. Our results are compatible with the best fit curves from Weigel et al. except for the massive end of the quiescent galaxies --where more complex phenomena need to be considered.
}
\end{figure*}

To generate our Figure \ref{fig:weigel} we take their best-fit values of the blue parameters
\begin{equation}\label{eq:blue_weigel}
\begin{split}
    \phi_{*b} & = 10^{-2.423}/h^3 Mpc^{-3}\\
    \alpha_{b} & = -1.21\\
    m_{*b} & = 10^{10.60}M_{\odot}
\end{split}
\end{equation}
and use them in equations \eqref{eq:schechter} and \eqref{eq:schechter_blue}. Then we generate the satellite and central curves according to equation \eqref{eq:nblue}. For simplicity we gather all of the prefactors in the blue amplitudes \eqref{eq:nblue_amplitude} into a single  parameter and use the relation between fraction of satellites and probability. Therefore, we can write
\begin{equation}\label{eq:nblue_amplitude_validation}
    \begin{split}
        B_{c*} & = B(1-f_{\rm sat}) \\
        B_{s*} & = B(1-f_{\rm sat}) \ln \left(\frac{1}{ 1-f_{\rm sat}}\right) \, .
    \end{split}
\end{equation}
We determine the value of the parameter $B$ by imposing the reasonable condition that the sum of the blue amplitudes \eqref{eq:nblue_amplitude_validation} should equal the total blue sample from \cite{Weigel_2016}.
At this point, one could fix the fraction of satellite galaxies to a simple constant. Nonetheless, to make it more realistic and mass dependent we take the entire sample from \cite{Weigel_2016}, split by centrals and satellite galaxies (their Figure 16) and calculate the fraction \eqref{eq:fsat}. This is a simple procedure, and one could use a more sophisticated method, but suffices for our validation purposes. Finally, the fraction of satellite-quenched galaxies needs fine tuning. In this case, we simply take a known value from the literature $f_{\rho} \simeq 0.5$ \citep{Birrer_2014}. 

With all these ingredients and equations \eqref{eq:nred}, we plot both the active and quiescent galaxy populations (Figure \ref{fig:weigel}).
Our results are highly compatible with Weigel et al.'s best-fit curves. As expected, the main discrepancy shows in the massive-end of the red sample where other complex phenomena such as galaxy-galaxy merging processes dominate. 

All in all, our simple model successfully connects the growth and quenching processes with the different galaxy populations and justifies the appearance of a double Schechter function for the quiescent galaxies: the satellite-quenched galaxies, a subset of the blue population, and the mass-quenched galaxies with a different faint-end slope parameter.
For a more realistic prescription one would need to track down the history of the dark matter halos and their galaxies, the merger trees, and their time evolution to precisely know the time- and mass-dependence of the fraction of satellites, as well as the time evolution of the fraction of satellite-quenched galaxies. This will be the scope of future work within the SkyPy collaboration.

\section{Conclusions}
\label{sec:conclusions}

In this paper we focused on the theoretical description of galaxy demographics. The physical scene sets in halos hosting galaxies growing in mass, the so-called active population. As they become more massive, these central galaxies are subjected to mass quenching and cease star formation. The survival can become satellite galaxies with a probability of being satellite quenched due to the increase in environmental density. Satellite galaxies that survive this phenomenon will continue to grow until eventually becoming mass quenched as they reach a critical mass. 

In this picture we classified the quiescent population into mass-quenched galaxies and satellite-quenched galaxies. We distinguished between active galaxies (centrals and satellites) and quiescent galaxies (mass-quenched and satellite-quenched), describing the galaxy demographics with a set of continuity equations that we solved analytically. Such equations invoke two quenching mechanisms that transform star-forming galaxies into quiescent objects: mass quenching and satellite quenching. In this paper we made the necessary specification and explicitly included satellite galaxies, completing the description of current published sets of such continuity equations, e.g. \cite{Peng_2010}. This allowed us to provide a more accurate description of galaxies demographics so critical for the generation of realistic simulations. 

From the analytical solutions, we showed that the combination of the two quenching mechanisms produces a double Schechter function for the quiescent population. We demonstrated that the satellite-quenched galaxies are indeed a subset of the active galaxies and that the mass-quenched galaxies have a different faint-end slope parameter. The connection between  quenching and  galaxy populations reduced significantly the parameter space of our simulations. Instead of nine Schechter parameters, the same samples can be drawn by fixing the three star-forming Schechter parameters plus the fraction of satellite galaxies and the fraction of satellite-quenched galaxies. We then derived the analytical time dependence of the amplitude of the Schechter function for the first time in the literature. Comparison with empirical models showed the parametrisation used by \cite{Herbel} appears to be a sensible model.

Then we validated our model against SDSS DR7 data using the best-fitting model for the blue Schechter parameters from \cite{Weigel_2016} in the redshift bin $0.02 < z < 0.06$. We split their samples into centrals and galaxies to obtain the fraction of satellite galaxies and used a fixed known value for the fraction of satellite-quenched galaxies. The main discrepancy in our comparison showed in the massive-end of the red sample, where galaxy-galaxy merging effects are believed to dominate. We leave for future work the modelling of a more complex scenario, including galaxy-galaxy merging. Another extension of this work will consider the time dependence of the characteristic mass $m_{*}$ \citep{Herbel}. 

Finally, our model will be included in the \verb!skypy.galaxies! module of the \verb!SkyPy! library \citep{Amara_2021, skypy_collaboration_2020_3755531} after the publication of this manuscritp. In addition, the sonification, or transformation of physical data via sound, is becoming increasingly important to make astronomy accessible for those who are visually impaired, and to enhance visualisations and convey information that visualisation alone cannot (c.f \cite{harrison2021audio}). In this work we also made our main plot available in sound format using the \verb!STRAUSS! software \citep{james_trayford_2021_5776280}. This will be found in the \verb!SkyPy! documentation page.

\section*{Acknowledgements}

We would like to acknowledge all of the insightful comments from our colleagues in the SkyPy Collaboration, specially I. Harrison, R. Rollins and N. Tessore. We also acknowledge J. Trayford for helping us sonify our results.

The preparation of this manuscript was made possible by a number of software packages: \verb!NumPy!, \verb!SciPy! \citep{Scipy_2020}, \verb!Astropy! \citep{Astropy_2018}, \verb!Matplotlib! \citep{Matplotlib_2007}, \verb!IPython/Jupyter! \citep{Jupyter_2007}. 
We also employed the \verb!STRAUSS! software \citep{james_trayford_2021_5776280} to sonify our main results.


\bibliographystyle{mnras}
\bibliography{main}

\begin{thebibliography}{}
\makeatletter
\relax
\def\mn@urlcharsother{\let\do\@makeother \do\$\do\&\do\#\do\^\do\_\do\%\do\~}
\def\mn@doi{\begingroup\mn@urlcharsother \@ifnextchar [ {\mn@doi@}
  {\mn@doi@[]}}
\def\mn@doi@[#1]#2{\def\@tempa{#1}\ifx\@tempa\@empty \href
  {http://dx.doi.org/#2} {doi:#2}\else \href {http://dx.doi.org/#2} {#1}\fi
  \endgroup}
\def\mn@eprint#1#2{\mn@eprint@#1:#2::\@nil}
\def\mn@eprint@arXiv#1{\href {http://arxiv.org/abs/#1} {{\tt arXiv:#1}}}
\def\mn@eprint@dblp#1{\href {http://dblp.uni-trier.de/rec/bibtex/#1.xml}
  {dblp:#1}}
\def\mn@eprint@#1:#2:#3:#4\@nil{\def\@tempa {#1}\def\@tempb {#2}\def\@tempc
  {#3}\ifx \@tempc \@empty \let \@tempc \@tempb \let \@tempb \@tempa \fi \ifx
  \@tempb \@empty \def\@tempb {arXiv}\fi \@ifundefined
  {mn@eprint@\@tempb}{\@tempb:\@tempc}{\expandafter \expandafter \csname
  mn@eprint@\@tempb\endcsname \expandafter{\@tempc}}}

\bibitem[\protect\citeauthoryear{Amara et~al.,}{Amara
  et~al.}{2021}]{Amara_2021}
Amara A.,  et~al., 2021, \mn@doi [Journal of Open Source Software]
  {10.21105/joss.03056}, 6, 3056

\bibitem[\protect\citeauthoryear{Baldry et~al.,}{Baldry
  et~al.}{2012}]{Baldry_2012}
Baldry I.~K.,  et~al., 2012, \mn@doi [Monthly Notices of the Royal Astronomical
  Society] {10.1111/j.1365-2966.2012.20340.x}, 421, 621

\bibitem[\protect\citeauthoryear{Balogh, Navarro  \& Morris}{Balogh
  et~al.}{2000}]{Balogh_2000}
Balogh M.~L.,  Navarro J.~F.,   Morris S.~L.,  2000, \mn@doi [The Astrophysical
  Journal] {10.1086/309323}, 540, 113

\bibitem[\protect\citeauthoryear{{Birrer}, {Lilly}, {Amara}, {Paranjape}  \&
  {Refregier}}{{Birrer} et~al.}{2014}]{Birrer_2014}
{Birrer} S.,  {Lilly} S.,  {Amara} A.,  {Paranjape} A.,   {Refregier} A.,
  2014, \mn@doi [\apj] {10.1088/0004-637X/793/1/12}, \href
  {https://ui.adsabs.harvard.edu/abs/2014ApJ...793...12B} {793, 12}

\bibitem[\protect\citeauthoryear{{Efstathiou}, {Ellis}  \&
  {Peterson}}{{Efstathiou} et~al.}{1988}]{Efstathiou_1988}
{Efstathiou} G.,  {Ellis} R.~S.,   {Peterson} B.~A.,  1988, \mn@doi [\mnras]
  {10.1093/mnras/232.2.431}, \href
  {https://ui.adsabs.harvard.edu/abs/1988MNRAS.232..431E} {232, 431}

\bibitem[\protect\citeauthoryear{Fabian}{Fabian}{2012}]{Fabian_2012}
Fabian A.,  2012, \mn@doi [Annual Review of Astronomy and Astrophysics]
  {10.1146/annurev-astro-081811-125521}, 50, 455

\bibitem[\protect\citeauthoryear{{Gunn} \& {Gott}}{{Gunn} \&
  {Gott}}{1972}]{GunnGott_1972}
{Gunn} J.~E.,  {Gott} J.~Richard I.,  1972, \mn@doi [\apj] {10.1086/151605},
  \href {https://ui.adsabs.harvard.edu/abs/1972ApJ...176....1G} {176, 1}

\bibitem[\protect\citeauthoryear{Harrison, Trayford, Harrison  \&
  Bonne}{Harrison et~al.}{2021}]{harrison2021audio}
Harrison C.,  Trayford J.,  Harrison L.,   Bonne N.,  2021, Audio Universe Tour
  of the Solar System: using sound to make the Universe more accessible
  (\mn@eprint {arXiv} {2112.02110})

\bibitem[\protect\citeauthoryear{{Hartley} et~al.,}{{Hartley}
  et~al.}{2013}]{Hartley_2013}
{Hartley} W.~G.,  et~al., 2013, \mn@doi [\mnras] {10.1093/mnras/stt383}, \href
  {https://ui.adsabs.harvard.edu/abs/2013MNRAS.431.3045H} {431, 3045}

\bibitem[\protect\citeauthoryear{{Herbel}, {Kacprzak}, {Amara}, {Refregier},
  {Bruderer}  \& {Nicola}}{{Herbel} et~al.}{2017}]{Herbel}
{Herbel} J.,  {Kacprzak} T.,  {Amara} A.,  {Refregier} A.,  {Bruderer} C.,
  {Nicola} A.,  2017, \mn@doi [\jcap] {10.1088/1475-7516/2017/08/035}, \href
  {https://ui.adsabs.harvard.edu/abs/2017JCAP...08..035H} {2017, 035}

\bibitem[\protect\citeauthoryear{{Hunter}}{{Hunter}}{2007}]{Matplotlib_2007}
{Hunter} J.~D.,  2007, \mn@doi [Computing in Science and Engineering]
  {10.1109/MCSE.2007.55}, \href
  {https://ui.adsabs.harvard.edu/abs/2007CSE.....9...90H} {9, 90}

\bibitem[\protect\citeauthoryear{{Ilbert} et~al.,}{{Ilbert}
  et~al.}{2013}]{Ilbert_2013}
{Ilbert} O.,  et~al., 2013, \mn@doi [\aap] {10.1051/0004-6361/201321100}, \href
  {https://ui.adsabs.harvard.edu/abs/2013A&A...556A..55I} {556, A55}

\bibitem[\protect\citeauthoryear{{Ivezi{\'c}} et~al.,}{{Ivezi{\'c}}
  et~al.}{2019}]{LSST2019}
{Ivezi{\'c}} {\v{Z}}.,  et~al., 2019, \mn@doi [The Astrophysical Journal]
  {10.3847/1538-4357/ab042c}, \href
  {http://adsabs.harvard.edu/abs/2019ApJ...873..111I} {873, 111}

\bibitem[\protect\citeauthoryear{{Knobel}, {Lilly}, {Woo}  \&
  {Kova{\v{c}}}}{{Knobel} et~al.}{2015}]{Knobel_2015}
{Knobel} C.,  {Lilly} S.~J.,  {Woo} J.,   {Kova{\v{c}}} K.,  2015, \mn@doi
  [\apj] {10.1088/0004-637X/800/1/24}, 800, 24

\bibitem[\protect\citeauthoryear{{LIGO Scientific Collaboration}}{{LIGO
  Scientific Collaboration}}{2015}]{LIGO2015}
{LIGO Scientific Collaboration} 2015, \mn@doi [Classical and Quantum Gravity]
  {10.1088/0264-9381/32/7/074001}, \href
  {https://ui.adsabs.harvard.edu/abs/2015CQGra..32g4001L} {32, 074001}

\bibitem[\protect\citeauthoryear{{Larson}, {Tinsley}  \& {Caldwell}}{{Larson}
  et~al.}{1980}]{Larson_1980}
{Larson} R.~B.,  {Tinsley} B.~M.,   {Caldwell} C.~N.,  1980, \mn@doi [\apj]
  {10.1086/157917}, \href
  {https://ui.adsabs.harvard.edu/abs/1980ApJ...237..692L} {237, 692}

\bibitem[\protect\citeauthoryear{{Laureijs} et~al.,}{{Laureijs}
  et~al.}{2011}]{Euclid2011}
{Laureijs} R.,  et~al., 2011, arXiv e-prints, \href
  {https://ui.adsabs.harvard.edu/abs/2011arXiv1110.3193L} {p. arXiv:1110.3193}

\bibitem[\protect\citeauthoryear{{Li} \& {White}}{{Li} \&
  {White}}{2009}]{LiWhite_2009}
{Li} C.,  {White} S. D.~M.,  2009, \mn@doi [\mnras]
  {10.1111/j.1365-2966.2009.15268.x}, \href
  {https://ui.adsabs.harvard.edu/abs/2009MNRAS.398.2177L} {398, 2177}

\bibitem[\protect\citeauthoryear{{Muzzin} et~al.,}{{Muzzin}
  et~al.}{2013}]{Muzzin_2013}
{Muzzin} A.,  et~al., 2013, \mn@doi [\apj] {10.1088/0004-637X/777/1/18}, \href
  {https://ui.adsabs.harvard.edu/abs/2013ApJ...777...18M} {777, 18}

\bibitem[\protect\citeauthoryear{{Peng} et~al.,}{{Peng}
  et~al.}{2010}]{Peng_2010}
{Peng} Y.-j.,  et~al., 2010, \mn@doi [\apj] {10.1088/0004-637X/721/1/193},
  \href {https://ui.adsabs.harvard.edu/abs/2010ApJ...721..193P} {721, 193}

\bibitem[\protect\citeauthoryear{{Peng}, {Lilly}, {Renzini}  \&
  {Carollo}}{{Peng} et~al.}{2012}]{Peng_2012}
{Peng} Y.-j.,  {Lilly} S.~J.,  {Renzini} A.,   {Carollo} M.,  2012, \mn@doi
  [\apj] {10.1088/0004-637X/757/1/4}, \href
  {https://ui.adsabs.harvard.edu/abs/2012ApJ...757....4P} {757, 4}

\bibitem[\protect\citeauthoryear{{Perez} \& {Granger}}{{Perez} \&
  {Granger}}{2007}]{Jupyter_2007}
{Perez} F.,  {Granger} B.~E.,  2007, \mn@doi [Computing in Science and
  Engineering] {10.1109/MCSE.2007.53}, \href
  {https://ui.adsabs.harvard.edu/abs/2007CSE.....9c..21P} {9, 21}

\bibitem[\protect\citeauthoryear{{Planck Collaboration}}{{Planck
  Collaboration}}{2020}]{Planck2020}
{Planck Collaboration} 2020, \mn@doi [Astronomy & Astrophysics]
  {10.1051/0004-6361/201833880}, \href
  {https://ui.adsabs.harvard.edu/abs/2020A&A...641A...1P} {641, A1}

\bibitem[\protect\citeauthoryear{{Pozzetti} et~al.,}{{Pozzetti}
  et~al.}{2010}]{Pozzetti_2010}
{Pozzetti} L.,  et~al., 2010, \mn@doi [\aap] {10.1051/0004-6361/200913020},
  \href {https://ui.adsabs.harvard.edu/abs/2010A&A...523A..13P} {523, A13}

\bibitem[\protect\citeauthoryear{Price-Whelan et~al.,}{Price-Whelan
  et~al.}{2018}]{Astropy_2018}
Price-Whelan A.~M.,  et~al., 2018, \mn@doi [The Astronomical Journal]
  {10.3847/1538-3881/aabc4f}, 156, 123

\bibitem[\protect\citeauthoryear{{Sandage}, {Tammann}  \& {Yahil}}{{Sandage}
  et~al.}{1979}]{Sandage_1979}
{Sandage} A.,  {Tammann} G.~A.,   {Yahil} A.,  1979, \mn@doi [\apj]
  {10.1086/157295}, \href
  {https://ui.adsabs.harvard.edu/abs/1979ApJ...232..352S} {232, 352}

\bibitem[\protect\citeauthoryear{{Schechter}}{{Schechter}}{1976}]{Schechter_1976}
{Schechter} P.,  1976, \mn@doi [\apj] {10.1086/154079}, \href
  {https://ui.adsabs.harvard.edu/abs/1976ApJ...203..297S} {203, 297}

\bibitem[\protect\citeauthoryear{{Schmidt}}{{Schmidt}}{1968}]{Schmidt_1968}
{Schmidt} M.,  1968, \mn@doi [\apj] {10.1086/149446}, \href
  {https://ui.adsabs.harvard.edu/abs/1968ApJ...151..393S} {151, 393}

\bibitem[\protect\citeauthoryear{{SkyPy Collaboration}}{{SkyPy
  Collaboration}}{2020}]{skypy_collaboration_2020_3755531}
{SkyPy Collaboration} 2020, SkyPy, \mn@doi{10.5281/zenodo.4071945}, \url
  {https://doi.org/10.5281/zenodo.4071945}

\bibitem[\protect\citeauthoryear{Trayford}{Trayford}{2021}]{james_trayford_2021_5776280}
Trayford J.,  2021, james-trayford/strauss: v0.1.0 Pre-release,
  \mn@doi{10.5281/zenodo.5776280}, \url
  {https://doi.org/10.5281/zenodo.5776280}

\bibitem[\protect\citeauthoryear{{Virtanen} et~al.,}{{Virtanen}
  et~al.}{2020}]{Scipy_2020}
{Virtanen} P.,  et~al., 2020, \mn@doi [Nature Methods]
  {10.1038/s41592-019-0686-2}, \href
  {https://ui.adsabs.harvard.edu/abs/2020NatMe..17..261V} {17, 261}

\bibitem[\protect\citeauthoryear{Weigel, Schawinski  \& Bruderer}{Weigel
  et~al.}{2016}]{Weigel_2016}
Weigel A.~K.,  Schawinski K.,   Bruderer C.,  2016, Monthly Notices of the
  Royal Astronomical Society, 459, 2150

\makeatother
\end{thebibliography}


\bsp	
\label{lastpage}
\end{document}